\documentclass[aps,twocolumn,prl]{revtex4}
\usepackage[utf8x]{inputenc}
\usepackage{graphicx}
\usepackage{latexsym}
\usepackage{amsmath}
\usepackage{amssymb}
\usepackage{psfrag}
\usepackage{bm}
\newcommand{\bee}{\begin{eqnarray}}
\newcommand{\eee}{\end{eqnarray}}
\newcommand{\ba}{\begin{array}}
\newcommand{\ea}{\end{array}}
\newcommand{\bc}{\begin{center}}
\newcommand{\ec}{\end{center}}
\newcommand{\bi}{\begin{itemize}}
\newcommand{\ei}{\end{itemize}}

\begin{document}

\title{Note: Brownian motion of colloidal particles of arbitrary shape}%
\author{Bogdan Cichocki}         
\affiliation{Institute of Theoretical Physics, Faculty of Physics, University of Warsaw, Pasteura 5,
  02-093 Warsaw, Poland}

\author{Maria L. Ekiel-Je\.zewska}\thanks{Corresponding author: mekiel@ippt.pan.pl}
 \affiliation{Institute of Fundamental Technological Research,
             Polish Academy of Sciences, Pawi\'nskiego 5B, 02-106 Warsaw, Poland}

\author{Eligiusz Wajnryb}
 \affiliation{Institute of Fundamental Technological Research,
             Polish Academy of Sciences, Pawi\'nskiego 5B, 02-106 Warsaw, Poland}

\date{\today}

\begin{abstract}
The analytical expressions for the time-dependent cross-correlations of the translational and rotational Brownian displacements of a particle with arbitrary shape are derived. The reference center is arbitrary, and the reference frame is such that the rotational-rotational diffusion tensor is diagonal. 
\end{abstract}

\maketitle
In general, the diffusion and mobility matrices of a particle depend on the choice of the reference center which is used to describe the rotational and translational motion. The requirement that the rotational-translational mobility matrix $\bm{\mu}^{rt}$ is symmetric defines uniquely the so-called center of mobility. 
In Ref.~\cite{CEW2015}, this center was used to analyze the Brownian motion of a particle with arbitrary shape, and the corresponding 
analytical expressions for the time-dependent cross-correlations of the translational and rotational Brownian displacements were derived. 

In this note, we derive the analogical analytical expressions for the cross-correlations of the translational and rotational displacements of an {\it arbitrary} reference center.
We denote the time-dependent position of this center as $\mathbf{R}(t)$, and we introduce 
three mutually perpendicular unit vectors $\mathbf{u}%
^{(p)}(t)$, $p=1,2,3$, to describe the particle orientation at time $t$ \cite{CEW2015,Kraft}. The translational and rotational Brownian displacements are, 
\bee
&&\Delta \mathbf{R}(t) = \mathbf{R}(t) - \mathbf{R}(0),\\
&&
\Delta \mathbf{u}(t)=\frac{1}{2}\sum_{p=1}^{3}\mathbf{u}%
^{(p)}(0)
\times \mathbf{u}^{(p)}(t).\label{defDeltau}
\eee

Their dynamical cross-correlations 
are evaluated as the averages $\langle ... \rangle_0$ of their products with respect to the particle positions and orientations, using the conditional probability which satisfies the Smoluchowski equation  \cite{Kampen,SE2}, 
with 
the diffusion matrix,
\bee
\bm{D}=\left[\!\!\ba{c}\bm{D}^{tt}\;\bm{D}^{tr}\\\bm{D}^{rt}\; \bm{D}^{rr}
\ea \!\!  \right],
\eee 
We adopt 
 the frame of reference in which $\bm{D}^{rr}$ is diagonal, i.e. $
D^{rr}_{\mu \nu}=D_{\mu} \delta_{\mu \nu},$ 
with 
$\mu, \nu=1,2,3$. 

When an arbitrary reference center is chosen, the tensor $\bm{D}^{rt}$
contains non-vanishing antisymmetric part.
In this case to derive 
the
rotational-translational correlations one can proceed as it is presented in
Sec.\!~VIII of Ref. \!\!\!\cite{CEW2015} with the following modification. 
The correlation $\left\langle \!\mathbf{u}^{(p)}\!(t)\Delta \mathbf{R}(t)\right\rangle_{0}$ is evaluated in the analogical way as in Eq.~\!(79) from Ref.~\cite{CEW2015}, but with another second rank tensor $\mathbf{A}^{(p)}$\!, i.e.
\bee &&
\mathbf{A}^{(p)}=\mathbf{u}^{(p)}\times \bm{D}^{rt}-\frac{1}{2}\mathbf{u}%
^{(p)}\mathbf{V},\label{nadV}
\eee
where the components of vector $\mathbf{V}$ are given as %
$V_{\alpha}\mathbf{=}\sum_{\mu,\nu=1}^{3}\epsilon _{\alpha\mu\nu}D_{\mu\nu}^{rt}.$

The resulting off-diagonal Cartesian components are,
\vspace{-0.5cm}
\begin{widetext}
\vspace{-0.7cm}%
\begin{eqnarray}
&&\left\langle \Delta  \mathbf{u}%
(t) \Delta \mathbf{R}(t)\right\rangle _{0,\alpha \beta }= 
-\left[ (D_{\alpha \beta }^{rt}+D_{\beta \alpha }^{rt})\left( \frac{1}{4}+%
\frac{D_{\beta }-D_{\gamma }}{4\Delta }\right) +(D_{\alpha \beta
}^{rt}-D_{\beta \alpha }^{rt})\left( \frac{1}{6}+\frac{D-D_{\alpha }}{2\Delta 
}\right) ~\right] \frac{\text{e}^{-f^{(-)}t}-\text{e}^{-f_{\gamma }^{(1)}t}}{%
f^{(-)}-f_{\gamma }^{(1)}} \nonumber \\
&&-\left[ (D_{\alpha \beta }^{rt}+D_{\beta \alpha }^{rt})\left( \frac{1}{4}-%
\frac{D_{\beta }-D_{\gamma }}{4\Delta }\right) +(D_{\alpha \beta
}^{rt}-D_{\beta \alpha }^{rt})\left( \frac{1}{6}-\frac{D-D_{\alpha }}{2\Delta 
}\right) ~\right] \frac{\text{e}^{-f^{(+)}t}-\text{e}^{-f_{\gamma }^{(1)}t}}{%
f^{(+)}-f_{\gamma }^{(1)}} \nonumber \\
&&-\frac{D_{\alpha \beta }^{rt}+D_{\beta \alpha }^{rt}}{4}~\left[ \frac{%
\text{e}^{-f_{\alpha }^{(2)}t}-\text{e}^{-f_{\beta }^{(1)}t}}{f_{\alpha
}^{(2)}-f_{\beta }^{(1)}}+\frac{\text{e}^{-f_{\alpha }^{(1)}t}-\text{e}%
^{-f_{\beta }^{(1)}t}}{f_{\alpha }^{(1)}-f_{\beta }^{(1)}}\right] 
-\frac{D_{\alpha \beta }^{rt}-D_{\beta \alpha }^{rt}}{2}~\left[ \frac{%
\text{e}^{-f_{\alpha }^{(2)}t}-\text{e}^{-f_{\beta }^{(1)}t}}{f_{\alpha
}^{(2)}-f_{\beta }^{(1)}}-\frac{1-\text{e}^{-f_{\gamma }^{(1)}t}}{%
3f_{\gamma }^{(1)}}\right],
\end{eqnarray}%
\vspace{-0.3cm}
\end{widetext}\vspace{-0.6cm}
with the convention that everywhere in this note $(\alpha,\beta,\gamma)$ is a permutation of $(1,2,3)$, and
\bee
\!\!\!&&f_{\alpha}^{(1)} =3D-D_{\alpha}, \hspace{0.5cm}f_{\alpha }^{(2)} =3(D_{\alpha }+D),\;\;\;\alpha =1,2,3,\hspace{0.6cm}\\
\!\!\!
\!\!\!
&&f^{(\pm)} =6D\pm 2\Delta , \label{f+}
\eee\bee
\!\!\!&&
D=\frac{1}{3}(D_{1}+D_{2}+D_{3}),\label{trrr}\\
\!\!
\!\!\!&&\Delta = \sqrt{%
D_{1}^{2}+D_{2}^{2}+D_{3}^{2}-D_{1}D_{2}-D_{1}D_{3}-D_{2}D_{3}}.\;\;\;\;
\eee

\newpage
The diagonal Cartesian components read:%
\begin{eqnarray}
\!\!\!&&\!\!\!\!\left\langle \Delta  \mathbf{u}(t) \Delta \mathbf{R}(t)%
\right\rangle _{0,\alpha \alpha }= 
-\frac{D_{\alpha \alpha }^{rt}\!-\!D_{\beta \beta }^{rt}}{8}~\frac{\text{e}%
^{-f_{\gamma }^{(2)}t}\!-\!\text{e}^{-f_{\gamma }^{(1)}t}}{D_{\gamma }}~\nonumber \\
\!\!\!&&\!\!\!\!-\frac{%
D_{\alpha \alpha }^{rt}\!-\!D_{\gamma \gamma }^{rt}}{8}~\frac{\text{e}%
^{-f_{\beta }^{(2)}t}\!-\!\text{e}^{-f_{\beta }^{(1)}t}}{D_{\beta }} 
+\frac{D_{\alpha \alpha }^{rt}\!+\!D_{\beta \beta }^{rt}}{2}~\text{e}%
^{-f_{\gamma }^{(1)}t}t
\nonumber\\
\!\!\!&&\!\!\!\!+\frac{D_{\alpha \alpha }^{rt}\!+\!D_{\gamma \gamma
}^{rt}}{2}~\text{e}^{-f_{\beta }^{(1)}t}t.
\end{eqnarray}%

To derive the translational-translational correlations, the derivative (54) from 
Ref. \cite{CEW2015} must be changed into
\bee
\hspace{-0.3cm}
&&\frac{d}{dt}\!\left\langle \Delta \mathbf{R}(t)\Delta\mathbf{R%
}(t)\right\rangle_0 \nonumber \\
\hspace{-0.3cm}&&=\!2\left\langle \bm{D}^{tt}(t)\right\rangle_0
+\left\langle \mathbf{V}(t)\Delta\mathbf{R}(t)\right\rangle_0
+\left\langle \Delta\mathbf{R}(t)\mathbf{V}(t)\right\rangle_0.\;\;
\eee

The two last terms can be then calculated by decomposing the vector $\mathbf{%
V}$, defined under Eq. \eqref{nadV},  in the frame $\mathbf{u}^{(p)}$, $p=1,2,3$, and using the results for $%
\left\langle \mathbf{u}^{(p)}(t)\Delta\mathbf{R}(t)\right\rangle
_{0}$ obtained by the method described above.

The 
off-diagonal Cartesian components with $\alpha \!\!\ne \!\!\beta$ read,%
\begin{eqnarray}
&&\frac{1}{2}\left\langle \Delta \mathbf{R}(t)\Delta \mathbf{%
R}(t)\right\rangle _{0,\alpha \beta }=\frac{1-\text{e}^{-f_{\gamma }^{(2)}t}}{%
f_{\gamma }^{(2)}}D_{\alpha \beta }^{tt} \nonumber \\
&&+\frac{1}{2}(D_{\beta \gamma }^{rt}-D_{\gamma \beta }^{rt})(3D_{
\gamma \alpha}^{rt}-D_{\alpha \gamma }^{rt})~I(f_{\gamma }^{(2)},f_{\alpha
}^{(1)};t) \nonumber \\
&&+\frac{1}{2}(D_{\alpha \gamma }^{rt}-D_{\gamma \alpha }^{rt})(3D_{\gamma \beta
}^{rt}-D_{\beta \gamma }^{rt})~I(f_{\gamma }^{(2)},f_{\beta }^{(1)};t)
\nonumber \\
&&-(D_{\alpha \beta }^{rt}-D_{\beta \alpha }^{rt})(D_{\alpha \alpha
}^{rt}-D_{\beta \beta }^{rt})~I(f_{\gamma }^{(2)},f_{\gamma }^{(1)};t),
\end{eqnarray}
where 
\bee
\hspace{-0.25cm}&&I(f_{1},f_{2};t)=\frac{1}{f_{1}f_{2}}+\frac{\text{e}^{-f_{1}t}}{%
f_{1}(f_{1}-f_{2})}-\frac{\text{e}^{-f_{2}t}}{f_{2}(f_{1}-f_{2})}.\hspace{1cm}
\eee
The diagonal Cartesian components are,
\begin{eqnarray}
\!\!\!\!\!\!\!\!\!&&\frac{1}{2}\left\langle \Delta \mathbf{R}(t)\Delta \mathbf{%
R}(t)\right\rangle _{0,\alpha \alpha }\!\!=\!\!W(t)+\sum_{\sigma=\pm}\frac{1-\text{e}^{-f^{(\sigma)}t}}{%
f^{(\sigma)}}D_{\alpha \alpha }^{tt(\sigma)}
\nonumber \\
\!\!\!\!\!\!\!\!\!&&+\sum_{\sigma=\pm} \sum\limits_{\beta =1}^{3}P^{(\sigma)}\!\left( \alpha ,\beta \right)
I(f^{(\sigma)},f_{\beta }^{(1)};t),
\end{eqnarray}%
where%
\bee
\hspace{-0.25cm}&&{D}^{tt(\pm )}_{\alpha \alpha }=\left( \!\frac{1}{2}%
\mp \frac{3}{4}\frac{D_{\alpha }\!-\!D}{\Delta }\!\right) \left(\!
D^{tt}_{\alpha \alpha }\!-\!\frac{1}{3}\text{Tr}\boldsymbol{D}^{tt}\!\right) \nonumber \\
\hspace{-0.25cm}
&&\pm\frac{%
D_{\beta }\!-\!D_{\gamma }}{4\Delta }\left( D^{tt}_{\beta \beta }\!-\!D^{tt}_{\gamma \gamma
}\right), 
\eee
and
\bee
\hspace{-0.25cm}&&P^{(\pm )}(\alpha ,\beta )=-(D_{\alpha \gamma }^{rt}-D_{\gamma \alpha
}^{rt})^{2}\left( \frac{1}{3}\mp \frac{D-D_{\gamma }}{\Delta }\right) \hspace{0.5cm}\nonumber \\
\hspace{-0.2cm}&&+\!\left[
(D_{\alpha \gamma }^{rt})^{2}\!-\!(D_{\gamma \alpha }^{rt})^{2}\right] \left(\frac{1}{%
2}\mp \frac{D_{\alpha }\!-\!D_{\beta }}{2\Delta }\!\right)\!, \mbox{ for } \alpha \!\neq \!\beta,\;\;\hspace{0.4cm}\\&&
P^{(\pm )}(\alpha ,\alpha )=-P^{(\pm )}(\beta ,\alpha )-P^{(\pm )}(\gamma,\alpha ),%
\eee

Since $\sum_{\alpha=1}^{3} D^{tt(\pm)}_{\alpha\alpha}\!=\!0$, it is easy to see that $W(t)$ is 1/6 of  the mean square Brownian displacement,
\bee
\hspace{-0.25cm}&& W(t)\!=\!\frac{1}{6}\left\langle \Delta \mathbf{R}(t)\cdot
\Delta \mathbf{R}(t)\right\rangle_0. \eee
Moreover, it follows from our calculation that
\bee
W(t)\!=\!D_{cm}t+\frac{1}{3}%
\!\sum\limits_{\alpha =1}^{3}\!\frac{(D_{\beta \gamma }^{rt}\!\!-\!D_{\gamma \beta
}^{rt})^{2}}{(D_{\beta }\!+\!D_{\gamma })^{2}}\!\left( 1\!-\!\text{e}^{-(D_{\beta
}+D_{\gamma })t}\right)\!\!, \;\;\;\label{W} \eee
where $\beta, \gamma \ne \alpha$ and $D_{cm}$ is the self-diffusion coefficient referring to the center of mobility, 
\bee
\hspace{-0.25cm}&&D_{cm}=\frac{1}{3}\left[ \text{Tr}\bm{D}^{tt}-\sum\limits_{\alpha =1}^{3}\frac{%
(D_{\beta \gamma }^{rt}-D_{\gamma \beta }^{rt})^{2}}{D_{\beta }+D_{\gamma }}%
\right].
\eee

The importance of the center of mobility follows from Eq.~\eqref{W} -- this is the only reference center for which the mean square displacement is a linear function of time.\cite{CEJW_2012trans} 

For completeness, we remind the rotational-rotational correlations \cite{CEW2015}, which do not depend on the choice of the reference center. 
The off-diagonal elements vanish, and
the diagonal elements, 
$\alpha \!=\!1,2,3,$ are given by
\begin{eqnarray}
\hspace{-0.7cm}&&\left\langle \Delta\mathbf{u}(t)\Delta\mathbf{u}(t)\right\rangle _{0,\alpha \alpha } \!=
\frac{1}{6}-\frac{%
3(D\!-\!D_{\alpha })\!+\!\Delta }{12\Delta }\text{e}^{-f^{(-)} t}
\nonumber \\
\hspace{-0.7cm}&&
-\frac{3(D_{\alpha }\!-\!D)\!+\!\Delta }{%
12\Delta }\text{e}^{-f^{(+)}t}\!
-\!\frac{1}{4}\text{e}^{-f_{\alpha}^{(2)}
t}
\!+\!\frac{1}{4}\text{e}^{-f_{\alpha}^{(1)}
t}.\;\;\label{rrii}
\end{eqnarray}

In this work, the analytical expressions for the cross-correlations are more complex that their analogs in Ref.~\cite{CEW2015}, but the advantage is that they apply to an {\it arbitrary} reference center. Therefore, they can be easily used to account for the experimental results, obtained for any reference center. An example of the comparison of the theoretical expressions derived here with the measurements (made in Ref.~\cite{Kraft}) can be found in Ref.~\cite{CEW2016comment}.\\

\small{M.L.E.-J. and E.W. were supported in part by the Polish National Science Centre under Grant No. 2012/05/B/ST8/03010. M.L.E.-J. benefited from the scientific activities of the COST Action MP1305.}


\begin{thebibliography}{99}
\bibitem{CEW2015}
B. Cichocki, M. L. Ekiel-Je\.zewska and E. Wajnryb, J. Chem. Phys. {\bf 142}, 214902 (2015).
\bibitem{Kraft}
D. J. Kraft, R. Wittkowski, B. ten Hagen, K. V. Edmond, D. J. Pine and H. L\"owen, Phys. Rev. E {\bf 88}, 050301 (2013). 
\bibitem{Kampen} 
N. van Kampen, {\it Stochastic Processes in Physics and Chemistry, 3rd Edition}, North-Holland, 2007.
\bibitem{SE2}
R. B. Jones and P. N. Pusey, Annu. Rev. Phys. Chem. {\bf 42}, 137 (1991).
\bibitem{CEJW_2012trans}
B. Cichocki, M. L. Ekiel-Je\.zewska and E. Wajnryb, {J. Chem. Phys.} {\bf 136}, 071102 (2012).
\bibitem{CEW2016comment}
B. Cichocki, M. L. Ekiel-Je\.zewska and E. Wajnryb, to be published.
\end{thebibliography}
\end{document}